\documentstyle[12pt,mont98_desy,amssymb,epsfig]{article}

\def\be{\begin{equation}}
\def\ee{\end{equation}}
\def\bea{\begin{eqnarray}}
\def\eea{\end{eqnarray}}

\begin{document}

\title{IV.~~UNRUH EFFECT, SPIN POLARISATION AND THE DERBENEV--KONDRATENKO
FORMALISM
\footnote{Extended version of a talk presented
at the 15th ICFA Advanced Beam Dynamics Workshop: ``Quantum
Aspects of Beam Physics'', Monterey, California, U.S.A., January 1998.
Also in DESY Report 98--096, September 1998.}
}

\author{D.P. BARBER}

\address{Deutsches Elektronen--Synchrotron, DESY, \\
 22603 Hamburg, Germany. \\E-mail: mpybar@mail.desy.de}

\maketitle
\abstracts{ The relationship between the level of spin polarisation
caused by Unruh radiation as calculated by  Bell and Leinaas 
and that obtained from the Derbenev--Kondratenko formalism is explained.}

\section{Introduction}
In 1986 in the course of investigating  quantum fluctuations
in accelerated reference frames and striving to assign spin temperatures,
Bell and Leinaas (BL) \cite{bl87} found that in a perfectly aligned,
azimuthally uniform, weak focussing electron storage ring, the electron 
polarisation antiparallel to the dipole field is given by the formula
\begin{eqnarray}
  P_{eq} &=& \frac{8}{5 \sqrt{3}}
  \frac{1 - \frac{f}{6}}{1 - \frac{f}{18} +
  \frac{13}{360} f^{2}} .
 \end{eqnarray}
where $ f = (g - 2) Q^{2}_{z}/ (Q^{2}_{z} - \nu^{2})$ and $\nu = a \gamma$
\footnote{The notation is the same as in Article I.}.

Over most of the energy range $P_{eq}$ is  $8/5\sqrt{3}$~ i.e. $92.4\%$.
But as one approaches the resonance point $Q_{z} = \nu$ from
below, the polarisation dips to $-17 \%$  and then rises through
zero at the resonance energy to reach $99.2 \%$ before levelling
off again at $92.4 \%$.

Such behaviour is not exhibited in the 
DKM  {\it formula} (Article I, Eq.~(36)) which is based on
a calculation of spin motion driven by synchrotron radiation emission
in the laboratory frame. Indeed, in a perfectly aligned flat storage
ring  $\partial{\hat{n}}/{\partial{\delta}}$ is zero and
the polarisation is $92.4\%$ independently of energy.
At the time, the BL result caused considerable surprise and bafflement in 
the accelerator community.

\section{The solution}
However, the BL effect can be accommodated within the DKM {\it formalism }
and we were able to provide a detailed treatment \cite{bm88}. 
The full story can be found in \cite{bm88, minn88}
so that here, owing to space limitations, I will be exceedingly brief.

BL were primarily concerned with the effect of vertical orbit fluctuations
driven by the background Unruh radiation \cite{unruh98}.
In the laboratory frame these fluctuations stem from the fact that
synchrotron radiation photons are emitted at a small angle of order
$1/\gamma$ with respect to the horizontal plane and thus cause the particles
to recoil vertically. This  must also be taken into account when
considering the change in the $\hat{n}$ axis under photon emission
(Article I, Eq.~(35))
and the DKM formula for the polarisation along $\hat{n}$ then becomes:
\begin{equation}
P_{dk} = - \frac{8}{5 \sqrt{3}}
\frac
{\oint ds  \left< \frac{1}{\mid \rho \mid ^{3}}
\left[ \hat{b} \cdot \hat{n} - \hat{b} \cdot \vec{d} -
\frac{1}{6} \hat{s} \cdot \vec{f} \right] \right>_s}
{ \oint ds \left< \frac{1}{\mid \rho \mid ^{3}}
\left[
1 - \frac{2}{9}\ (\hat{n} \cdot \hat{s})^{2} + \frac{11}{18}\mid
\vec{d}\mid  ^2
- \frac{1}{18} \frac{\dot{\hat{s}}}{\mid \dot{\hat{s}} \mid}
\cdot (\hat{n} \times \vec{f}) + \frac{13}{360} \mid
\vec{f} \mid ^{2} \right] \right>_s}
\end{equation}
where the vector $\vec{f} \equiv - (2/\gamma)
\partial \hat{n}/\partial \beta_{z}$ and $\vec{d}=
\partial\hat{n}/\partial \delta$. See \cite{bm88,minn88} for notation
\footnote{In particular the vector $\vec{f}$ in Eq.~(2) and the quantity $f$
in Eq.~(1) are distinct.}.

If  $\partial \hat{n}/\partial\delta$ is zero as in the BL ring, 
the terms containing the very small quantity $\vec{f}$ come into play.
Then we obtain:
\begin{eqnarray}
P_{dk} &=& - \frac{8}{5 \sqrt{3}} \frac{1 - \frac{F}{6}}
{1 - \frac{F}{18} + \frac{13}{360}\ F^{2}}
 \end{eqnarray}
where $ F = \frac{2}{\gamma} + f$. 

Thus we have recovered the BL result except for the extra piece
$2/\gamma$. Near to the resonance this is negligible compared to
the resonance term and so near to the resonance we may consider
the two results to be in agreement. Thus the vertical kicks imparted to 
the orbit by the Unruh radiation of BL have been identified
with vertical recoils caused by synchrotron radiation. 

Further instructive interpretations of synchrotron radiation can be found in
\cite{mane90,lieu95}. In  \cite{lieu95} synchrotron 
radiation emission is considered to result from  `inverse Compton scattering'
of electrons from the virtual photons of the deflecting magnetic field and 
the spin dependent Compton cross--section is used to obtain the radiation
distribution. It would be interesting to see if an extension of this 
calculation emphasising spin effects could simulate the Sokolov--Ternov effect.

\section*{Acknowledgments}
I thank S. R. Mane for fruitful collaboration on this and other aspects
of spin polarisation in storage rings. I would like to thank J.D. Jackson,
W. G. Unruh and H. Rosu for useful exchanges of ideas.

\end{document}